# Distinguishing electronic and vibronic coherence in 2D spectra by their temperature dependence

*Václav Perlík,[†] Craig Lincoln,[‡] František Šanda[†] and Jürgen Hauer[‡]*

[†]Institute of Physics, Faculty of Mathematics and Physics, Charles University, Ke Karlovu 5, Prague 121 16, Czech Republic

[‡] Photonics Institute, Vienna University of Technology, Gusshausstrasse 27, 1040 Vienna, Austria

Long-lived oscillations in 2D spectra of chlorophylls are at the heart of an ongoing debate. Their physical origin is either a multi-pigment effect, such as excitonic coherence, or primarily stems from localized vibrations. In the present work, we analyze distinguishing characteristics of relative phase difference measured between diagonal- and cross-peak oscillations. While direct discrimination between the two scenarios is obscured when peaks overlap, their sensitivity to temperature provides a stronger argument. We show that vibrational (vibronic) oscillations change relative phase with temperature, while electronic oscillations are only weakly dependent. This highlights that studies of relative phase difference as a function of temperature provide a clear and easily accessible method to distinguish between vibrational and electronic coherences.

**KEYWORDS** Two-dimensional electronic spectroscopy, long lived coherences, quantum effects in biology, molecular dimers, vibronic dimers .

In the early stage of photosynthesis, excitation is created by absorption of a photon, followed by exciton migration from the peripheral antenna complexes to the reaction center with remarkably high efficiency.[1] Expansion of 2D spectroscopy[2] to the visible domain provided fresh impetus to photosynthesis research when Engel et al.[3] reported oscillations in the 2D spectra of the Fenna-Mathews-Olson (FMO) complex of bacteriochlorophylls. These oscillations survive over a picosecond and were ascribed to beating between delocalized excitonic (electronic) states.[3] The picosecond lifetime is unexpected for electronic coherence and despite attempts at explanation by adopting less common relaxation theories,[4] such a longevity sparked a still ongoing debate about the nature of the underlying coherence phenomena. Engel et al.[3] speculated that the surrounding protein shields the electronic coherence from dephasing interaction with the solvent environment. Low-frequency vibrations on individual pigments, the alternative readily explaining the observed lifetimes, were ruled out because of the observed out-of-phase (180°) difference between the diagonal- and anti-diagonal peak widths as a function of population time.

Nonetheless, the original argument triggered a critical discussion,[5-14] as it quickly became understood that the out-of-phase oscillatory pattern of diagonal and cross peaks cannot serve as a strong criterion for identification of electronic coherence. For example, Christensson et al.[5] showed that it depends strongly on excitation conditions, i.e. the overlap between absorption and excitation spectrum. Cheng and Fleming[6] showed that electronic and vibrational coherences can be distinguished by comparison of the rephasing (R) and non-rephasing (NR) portions of a 2D signal; where electronic beatings will only appear as oscillations on the NR diagonal-peaks and on R cross-peaks. Contrastingly, vibrational coherences do not show similar specificity and oscillations can be present for both NR and R signals at any type of peak (Figure 1). Experiment supported this view.[7] Other studies ruled out the purely electronic Frenkel excitons to be consistent with all observed features of FMO.[8] There are two concerns about Chengs' approach:[9-10] i) Purely rephasing and non-rephasing spectra can only be obtained with delta-pulses, i.e., with finite pulses the two signals are, to some extent, always mixed. ii) The current state of discussion accepts the important role of vibrations, even in primarily electronic scenarios, as they can support long-lived electronic coherences in FMO by means of vibronic dynamics.[11-12] Formally, the argument by Cheng and Fleming is based on comparing a four-level system (4LS) of vibrational character with a three-level system (3LS) of electronic character. In vibronic systems, electronic coupling can redistribute dipole strength to enhance excited state over ground state coherences[13] or vice versa.[14] This results in a 3LS-response pattern, which can explain all experimental findings in FMO,[13] without the need to invoke long-lived electronic coherences.

To follow recent discussions electronic (E), vibrational (V), and vibronic (P) model dynamics shall be specified. Constituting chromophores ($i = 1,2$) of the electronic dimer (E) are treated as two-level electronic systems with transition frequency $\varepsilon^i$ between ground $|g^i\rangle$ and excited $|e^i\rangle$ electronic states $\widehat{H}_e^i = \hbar\varepsilon^i|e^i\rangle\langle e^i|$. We next adopt many-body notation by introducing creation operators of exciton $\hat{A}^{i\dagger} = |e^i\rangle\langle g^i|$ and their annihilation adjoints $\hat{A}^i$. The two chromophores are coupled by resonant (dipole-dipole) interaction $\hbar J$. In the Heitler-London approximation the excitation of one chromophore is accompanied by de-excitation of the other, resulting in dimer hamiltonian

$$\widehat{H}_E = \sum_{i=1}^{2}\widehat{H}_e^i + \hbar J\left(\hat{A}^{1\dagger}\hat{A}^2 + \hat{A}^{2\dagger}\hat{A}^1\right) + \hbar\Upsilon\hat{A}^{1\dagger}\hat{A}^{2\dagger}\hat{A}^1\hat{A}^2, \qquad (1)$$

where $\Upsilon$ is a double-excitation frequency shift.

Underdamped vibration shall be modelled by a harmonic oscillator of frequency $\omega^i$ with displacement $d^i$ between ground and excited state potential surfaces of chromophore

$$\widehat{H}_v^i = \left[\frac{\hat{p}^{i\,2}}{2m} + \frac{1}{2}m\omega^{i\,2}\hat{q}^{i\,2}\right]|g^i\rangle\langle g^i| + \left[\frac{\hat{p}^{i\,2}}{2m} + \frac{1}{2}m\omega^{i\,2}(\hat{q}^i - d^i)^2\right]|e^i\rangle\langle e^i|\,; \qquad (2)$$

The simplest model of vibrational dynamics (V) assumes a single chromophore modulated by a vibration $\widehat{H}_V = \widehat{H}_e^1 + \widehat{H}_v^1$.

Interplay between resonant coupling in dimer (Eq. (1)) and underdamped on-site vibrations (Eq. (2)) induce vibronic dynamics (P)

$$\widehat{H}_P = \widehat{H}_E + \sum_{i=1}^{2}\widehat{H}_v^i \qquad (3)$$

Vibrations of Eq (2) are damped in Landau-Teller relaxation[15], i.e. coupled to a local bath[16] of spectral density $\zeta(\omega) = \frac{\lambda\Lambda\omega}{\omega^2+\Lambda^2}$ where $\Lambda$ is relaxation rate and $\lambda$ reorganization energy. Electronic dimer (E) is similarly suplemented with local spectral diffusion (see Appendix A2) .

The dynamics of 2D spectra is a combination of population transfer and lineshape effects (pure dephasing), induced by bath fluctuations of Hamiltonian's eigenstates and eigenvalues, respectively[17], are treated in the spirit of Zhang et al.[18] The population transport between eigenstates is thus decribed by quantum master equation. Line-shape effects are accounted for using second cumulant[17]. Calculations are performed at finite temperature $T$, taking into account detailed balance for the master equation and the fluctuation-dissipation theorem for cumulants (for more details see Appendix A3).

We start discussion of oscillatory components of 2D with qualitative diagrammatic analysis[19] for ground state bleach (GSB) and stimulated emission (SE). Interference from the ESA-pathways[20] are neglected for simplicity by formally taking large two exciton frequency shift $\Upsilon \to \infty$.

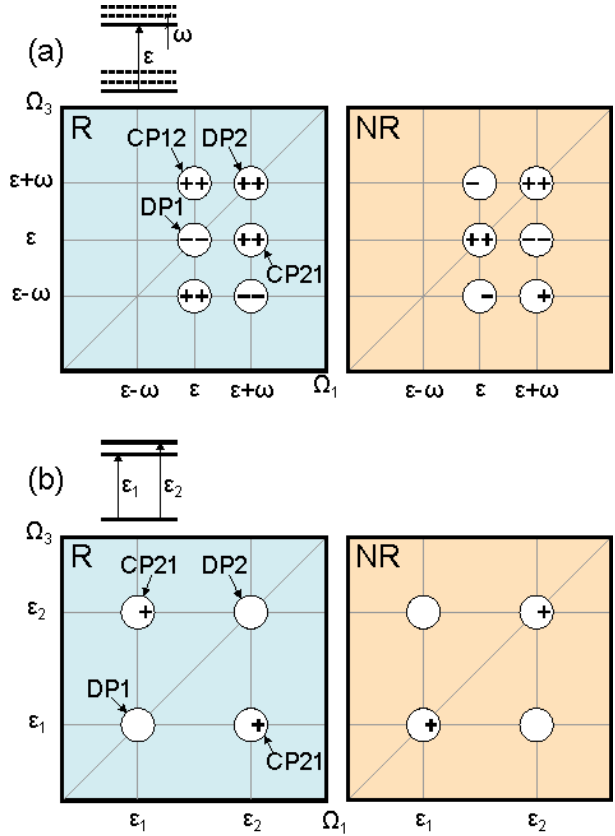

Figure 1(a) Schematic 2D-map of oscillatory phases for vibrational system (Eg (5)). Part of spectra related to first two vibrational levels over ground and over excited states. Signals oscillate as $A \mp B \cos \omega t_2, B > 0$, the relevant sign is indicated at left side of circles for GSB and right for SE, no sign indicate vanishing oscillations $B = 0$. R and NR denote rephasing and non-rephasing spectra, respectively. (b) Peaks structure and oscillatory phases of the form $A \mp B \cos([\varepsilon_2 - \varepsilon_1]t_2), B > 0$ for electronic dimer (Eq(6)).

Vibrational and electronic 2D peak structure is compared in Fig 1. We show two lowest vibrational peaks (and related cross-peaks), i.e. Figure 1a refers to four level systems of Refs,[6, 19] whereas the dimer in Figure 1b comprises a unique ground state and two electronic excitons (1). 2D peaks red-detuned from the lowest frequency of absorption (1D) require the existence of an energy level above the electronic ground state (Figure 1a) and serve as evidence of a ground state vibrational level.

We next demonstrate the capacity of the integrated pump-probe (PP) to discriminate between E and V dynamics suggested by Ref.[21] Strong out-of-phase oscillatory features on the vibrational lower diagonal (DP1) peak and cross peak CP21 (Figure 1a) are resolved in 2D but cancel when the signal is integrated over $\Omega_1$ (i.e., a PP-signal). In contrast, the electronic coherence of Figure 1b reveals in-phase oscillations which are retained after $\Omega_1$-integration. In short, comparison of PP and 2D-signals can serve as means to separate electronic and vibrational coherences in agreement with the wavepacket formalism of Yuen-Zhou et al.[21]

The diagrammatic framework[19] underlying the phases in Figure 1, results always in predictions of in/out of phase oscillations. More general phase shifts of the form $A + B\,cos(\omega t_2 + \phi)$, can only be explained by bypassing the assumptions behind the methodology used by Butkus et al.[19] Avoiding the secular form of transport equations[22] and invoking non-standard population to coherence transport was shown to allow for arbitrary phase shifts. The $\phi_{D2} - \phi_{CP21} = 90°$ phase shift, observed in FMO, was interpreted as evidence for such a "quantum transport",[23] i.e. it was argued that it indicates dynamical connection between population and coherence elements of the density matrix. We hereafter examine alternative explanations for a 90° phase difference. The analysis by Panitchayangkoon et al.[23] disregards effects from peak overlap, but as we will show, phase differences heavily depend on such overlaps.

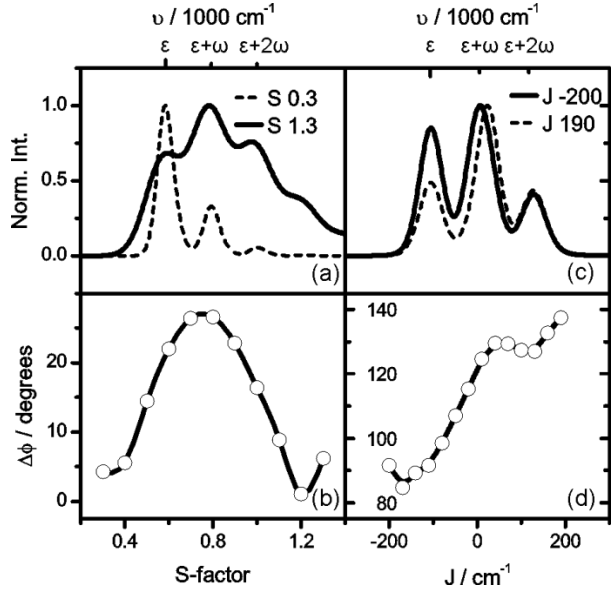

Figure 2 Left panels: Vibrational modulation of a single chromophore (a) Absorption spectra and (b) relative phase ($\Delta\phi$) between DP2 and CP21 peaks of 2D rephasing spectra as a function of Huang-Rhys factor $S$. Parameters: $\varepsilon = 16000$ cm$^{-1}$ and $\omega = 14000$ cm$^{-1}$

Right panels: Vibronic dimer near resonance $\varepsilon_1 \cong \varepsilon_2 + \omega$. (c) Absorption spectrum and (d) relative phase ($\Delta\phi$) between DP2 and CP21 peaks of rephasing spectra as a function of resonant coupling $J$. Dipoles of both chromophores are parallel and of the same value. Bath parameters $T = 300$ K, $\lambda = 600$ cm$^{-1}$ and $\Lambda = 100$ cm$^{-1}$ are same for all panels.

In Figure 2ab, we demonstrate emergence of phase shifts (details of the phase retrieval procedure can be found in the Appendix A1) when overlapping peaks of vibrational modulation. Overlaps were controlled by the Huang-Rhys factor $S = \frac{d^2 m\omega}{2\hbar}$, which scales linearly with peak width for Landau-Teller relaxation dynamics. For small overlaps ($S \to 0$), the 2D-maps depicted in Figure 1 are reproduced, i.e., $\Delta\phi \to 0$. Increasing $S$ broadens peaks and redistributes dipole strength to higher vibrational levels (Figure 2a). Spectral overlap between adjacent absorption peaks results in complex interferences, which significantly affect $\Delta\phi$ (Figure 2b) and peaks DP2 and CP21 no longer oscillate in phase. Phase varies significantly between 0° and appr. 30°, however we have not been able to reproduce 90° shifts reported in FMO.[23]

Phase variations are enhanced when introducing vibronic dynamics (Figures 2c,d). Dimer (Eq. (3)) was parameterized as a donor-acceptor system, where the acceptor's transition frequency is red shifted by about one vibrational quantum $\varepsilon_1 \cong \varepsilon_2 + \omega$. Thus vibrational ground at $|e_1\rangle$ and first excited level at $|e_2\rangle$ are resonant, covered by one peak in the absorption spectrum. Increasing coupling $J$ redistributes strength of transition dipoles and affects the retrieved values of $\Delta\phi$. Variation of $J$ allow to obtain arbitrary values of $\Delta\phi$, including $\Delta\phi = 90$ °, without the need to invoke the special transport dynamics of Ref [23]. $\Delta\phi$ deviating from 0° or 180° is thus not indicative for electronic coherence, as peak overlaps in vibronic systems yields similar effects.

Strong variations of phases observed in Figure 2 result from overlapping peaks DP1 and CP21 carrying opposite phases (Fig 1a). In contrast, all peaks of the electronic case oscillate in-phase (Fig 1b). This suggests, for an electronic system, independence of $\Delta\phi$ on absorption peak overlap, namely for the total (R + NR) 2D-signal. Similar ideas underlie recent observations that vibrational and electronic coherence show distinctly different behavior upon increasing disorder.[24] Here, we refine this idea into a readily accessible experimental variable to distinguish the two scenarios, namely temperature $T$.

In Figure 3 we compare temperature dependence of absorption spectra and phase shift for electronic (Figure 3a,b) and vibronic (Figure 3c,d) dimer. Chromophores of V model (Eq.(3)) were chosen symmetric $\varepsilon_1 = \varepsilon_2$, vibrational frequency correspond to energy splitting of E dimer. Vibrational structure was restricted to the two lowest levels and parameterized to yield similar absorption spectra at low temperatures.

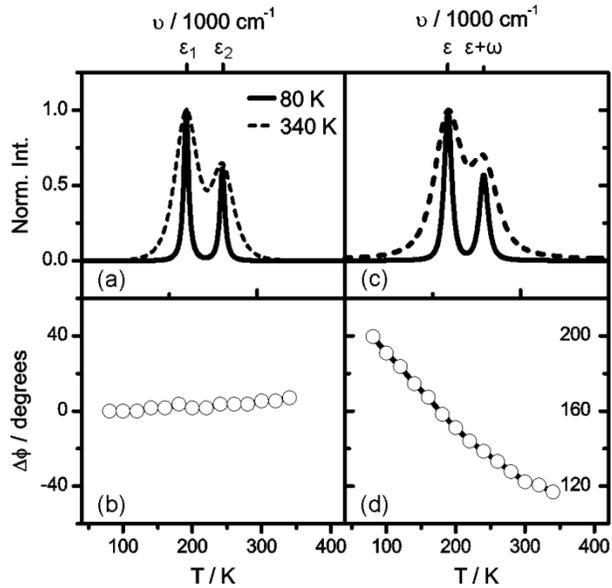

Figure 3. Temperature ($T$) dependent absorption spectra and $\Delta\phi$ for (a,b) electronic and (c,d) vibronic systems. Simulations are based on the total 2D-signal, i.e. R+NR. Parameters: Electronic: $\varepsilon_1 = 16000$ cm$^{-1}$, $\varepsilon_2 = 16200$ cm$^{-1}$, $J = -50$ cm$^{-1}$, $\lambda = 25$ cm$^{-1}$, $\Lambda = 80$ cm$^{-1}$. Vibronic: $\varepsilon_1 = \varepsilon_2 = 16000$ cm$^{-1}$, $\omega = 200$ cm$^{-1}$, $J = -30$ cm$^{-1}$ and $\lambda = 20$ cm$^{-1}$, $\Lambda = 40$ cm$^{-1}$, $S = 0.8$.

A rise in temperature broadens absorption peaks and increases overlap similarly in Fig 3a, 3c. Phase shift $\Delta\phi$, on the other hand, differs a lot. It is only weakly dependent on $T$ for electronic dimer (Fig 3b), but it is significantly stronger for the vibronic system (Fig. 3d). We analyzed the total (R+NR) 2D-signal to avoid experimental ambiguities related to pulse overlap effects present in pure R (NR) spectrum, however, their analysis would yield similar conclusions.

In conlusion, we have shown that the relative phase of peak oscillation in 2D maps is sensitive to peak overlap. In the case of vibrational origin of coherences, resulting values of $\Delta\phi$ vary between 0° and 180°. Contrastingly, purely electronic coherence leads to a $\Delta\phi$ that is insensitive to peak broadening effects (Figures 3a,b). Thus, in 2D-spectroscopy, electronic and vibrational (vibronic) coherences can be distinguished by studying the temperature dependence of their relative phase difference ($\Delta\phi$). The theoretical predictions made in this work can be readily verified by 2D-measurements at different temperatures and which have been reported.[25]

Appendix A1

Precise definition of phase meets several problems, even when 2D-shapes are well resolved. Oscillatory phase depends strongly on the precise position in the 2D-map[24], and retrieving phase from fixed $(\omega_1, \omega_3)$ point becomes further obscured by time-dependent Stokes shift. More meaningful results are obtained when the position $(\omega_1, \omega_3)$ with maximal amplitude of signal over the peak is identified for each position along waiting time $t_2$. This procedure, inter alia, correctly reproduces Figure 1 in the limit of non-overlapping peaks. The phase shift $\Delta\phi$ is finally obtained by taking the amplitude oscillations of DP2 and CP21 peaks into the Fourier domain with respect to waiting time $t_2$. Both peaks oscillate at the same frequency, i.e. the strongest Fourier component was always found at the same frequency. The phase shift $\Delta\phi$ is retrieved by comparing imaginary parts of these strongest Fourier components.

Appendix A2

Landau-Teller relaxation is induced by a coupling to a local bath of harmonic oscillators ($\Omega_k^i$ is a frequency of a $k$-th bath mode, $\hat{B}_k^{i\dagger}/\hat{B}_k^i$ are its creation/annihilation operators)

$$\hat{H}_b^i = \hbar \sum_k \Omega_k^i \left( \kappa_k^i \hat{B}_k^i \, \hat{b}^{i\dagger} + \kappa_k^i \hat{B}_k^{i\dagger} \hat{b}^i \right) + \hbar \sum_k \Omega_k^i \hat{B}_k^{i\dagger} \hat{B}_k^i \, ; \qquad \hat{b}^{i\dagger} = \sqrt{\frac{m\omega_i}{2\hbar}} \left( \hat{q}^i - \frac{i}{m\omega_i} \hat{p}^i \right)$$

with spectral density $\zeta(\omega) \equiv \sum_k {\Omega_k^i}^2 \kappa_k^i \kappa_k^i \delta(\omega - \Omega_k^i)$. The quantum transport master equation shall account for off-diagonal part of coupling and in eigenbasis of $\hat{H}_V$ ($\hat{H}_P$) reads

$$\frac{d}{dt}\rho_{\nu\mu}(t) = -\frac{i}{\hbar}(\varepsilon_\nu - \varepsilon_\mu)\rho_{\nu\mu}(t) + \sum_{\delta\beta} R_{\nu\mu,\delta\beta}\rho_{\delta\beta}(t),$$

where the relaxation kernel is taken in secular form

$$R_{\nu\mu,\delta\beta} = \sum_i \left( \sum_\gamma \delta_{\nu\delta}(n_{\gamma\beta} c^i_{\mu\gamma} c^i_{\beta\gamma} \zeta_{\gamma\beta} + (1+n_{\beta\gamma}) c^i_{\gamma\mu} c^i_{\gamma\beta} \zeta_{\beta\gamma}) \right.$$

$$+ \sum_\gamma \delta_{\beta\mu}(n_{\gamma\delta} c^i_{\nu\gamma} c^i_{\delta\gamma} \zeta_{\gamma\delta} + (1+n_{\delta\gamma}) c^i_{\gamma\nu} c^i_{\gamma\delta} \zeta_{\delta\gamma}) - n_{\mu\beta} c^i_{\delta\nu} c^i_{\beta\mu} \zeta_{\mu\beta}$$

$$\left. - (1+n_{\beta\mu}) c^i_{\nu\delta} c^i_{\mu\beta} \zeta_{\beta\mu} - n_{\nu\delta} c^i_{\delta\nu} c^i_{\beta\mu} \zeta_{\nu\delta} - (1+n_{\delta\nu}) c^i_{\nu\delta} c^i_{\mu\beta} \zeta_{\delta\nu} \right)$$

where $\zeta_{\alpha\beta} \equiv \zeta(\varepsilon_\alpha - \varepsilon_\beta)$ and $n_{\alpha\beta} \equiv \dfrac{1}{e^{\hbar(\varepsilon_\alpha - \varepsilon_\beta)/k_B T} - 1}$ and $c^i_{\delta\nu} = \langle \delta | \hat{b}^i | \nu \rangle (1 - \delta_{\nu\beta})$.

Relaxation of electronic dimer is induced by spectral diffusion Hamiltonian $\hat{H}^i_{b_o} = \sum_k \Omega^i_k \kappa^i_k \hat{B}^{i\dagger}_k \hat{A}^{i\dagger} \hat{A}^i +$ h.c.$+\hbar k \Omega ki B ki \dagger B ki$ and treated similarly in eigenbasis of $H_E$ with $c\delta\nu i = \delta | Ai\dagger Ai | \nu(1-\delta\nu\beta)$.

Appendix A3

In the absence of ESA, the 2D-signal comprises four R and four NR Liouville space pathways.[18] We illustrate their evaluation on R-pathways.

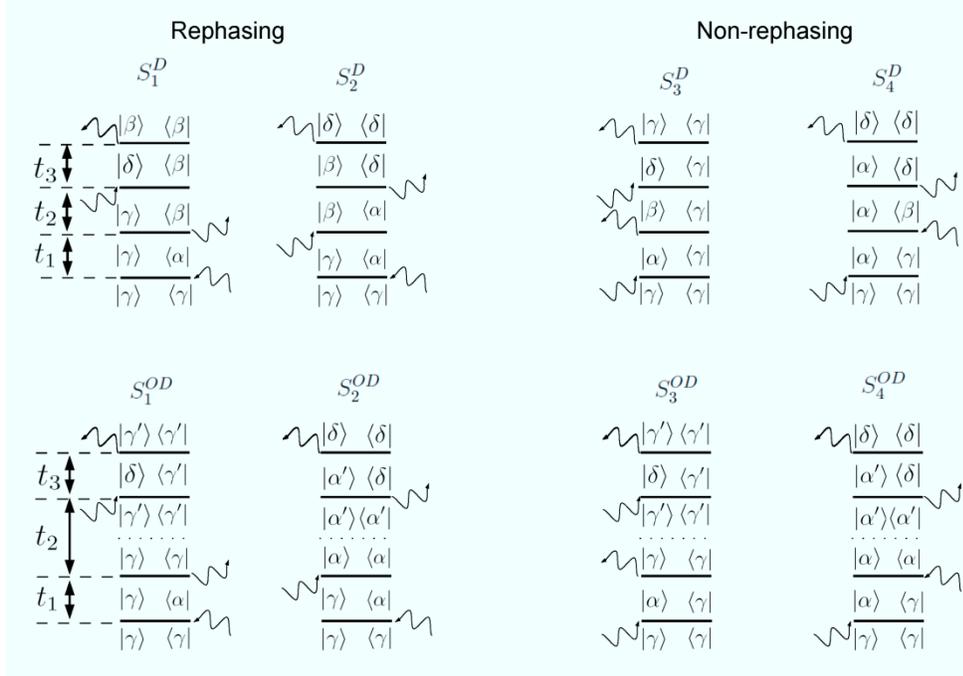

Top row diagrams keeps the exciton-index unchanged (in $\hat{H}_S$ eigenbasis) during $t_2$ evolution, e.g. stimulated emission in rephasing signal

$$S_2^D(t_1,t_2,t_3) = \sum_{\alpha\beta\delta} \mu_{\delta\beta} G_{\beta\delta,\beta\delta}(t_3)\, \mu_{\delta\alpha} G_{\beta\alpha,\beta\alpha}(t_2) \mu_{\beta\gamma} G_{\gamma\alpha,\gamma\alpha}(t_1) \mu_{\alpha\gamma} \rho_{\gamma\gamma} L_2^{\alpha\delta\beta}(t_1,t_2,t_3)$$

The other pathways undergo a quantum jump during $t_2$

$$S_2^{OD}(t_1,t_2,t_3) = \sum_{\alpha\delta\alpha'\neq\alpha} \mu_{\delta\alpha'} G_{\alpha'\delta,\alpha'\delta}(t_3)\, \mu_{\delta\alpha'} G_{\alpha'\alpha',\alpha\alpha}(t_2) \mu_{\alpha\gamma} G_{\gamma\alpha,\gamma\alpha}(t_1) \mu_{\alpha\gamma} \rho_{\gamma\gamma} \tilde{L}_2^{\alpha\alpha'\beta}(t_1,t_2,t_3)$$

Their weights are dictated by Green function solution $G$ to transport master equation.[18]

The $L_2^{\alpha\beta\delta}$ stands for phase accumulated during evolution by diagonal fluctuation $V_\alpha(t) = \hbar \sum_{k,i}(\hat{B}_k e^{-i\Omega_k t} + \hat{B}_k^\dagger e^{i\Omega_k t}) c_{\alpha\alpha}^i \kappa_k$ and can be readily defined for the first diagram

$$L_2^{\alpha\beta\delta} = \langle e_{-}^{\frac{i}{\hbar}\int_0^{t_1+t_2} d\tau V_\alpha(\tau)}\, e_{-}^{\frac{i}{\hbar}\int_{t_1+t_2}^{t_1+t_2+t_3} d\tau V_\delta(\tau)}\, e_{+}^{-\frac{i}{\hbar}\int_{t_1}^{t_1+t_2+t_3} d\tau V_\beta(\tau)} \rangle.$$

The peculiar point is the defintion for the other diagram, since the quantum jump can appear at any time during $t_2$. We approximated by assuming that the jump appeared at the middle point.

$$L_2^{\alpha\beta\delta} = \langle e_-^{\frac{i}{\hbar}\int_0^{t_1+t_2/2} d\tau V_\alpha(\tau)} e_-^{\frac{i}{\hbar}\int_{t_1+t_2/2}^{t_1+t_2} d\tau V_{\alpha\prime}(\tau)} e_-^{\frac{i}{\hbar}\int_{t_1+t_2}^{t_1+t_2+t_3} d\tau V_\delta(\tau)} e_+^{-\frac{i}{\hbar}\int_{t_1+t_2/2}^{t_1+t_2+t_3} d\tau V_{\alpha\prime}(\tau)} e_+^{-\frac{i}{\hbar}\int_{t_1}^{t_1+t_2/2} d\tau V_\alpha(\tau)} \rangle$$

These integrations can be readily carried out exactly by second cumulant.